\documentclass[prd,twocolumn,nofootinbib,english]{revtex4}
\usepackage{amsmath}
\usepackage{amsfonts,color}
\usepackage{amssymb,float}
\usepackage[utf8]{inputenc}
\usepackage{color}
\usepackage[unicode=true,pdfusetitle,
 bookmarks=true,bookmarksnumbered=true,bookmarksopen=true,bookmarksopenlevel=3,
 breaklinks=false,pdfborder={0 0 1},backref=false,colorlinks=true,citecolor    = blue]
 {hyperref}
\usepackage{enumitem}

\usepackage{etoolbox}
\usepackage{graphicx}
\usepackage{appendix}
\usepackage{hyperref}
\usepackage{url}
\usepackage{dirtytalk}
\usepackage[dvipsnames]{xcolor}
\colorlet{Mycolor1}{green!10!orange!90!}
\usepackage{array}
\newcolumntype{P}[1]{>{\centering\arraybackslash}p{#1}}
\newcolumntype{M}[1]{>{\centering\arraybackslash}m{#1}}
\begin{document}

\title{On the degeneracy between $f\sigma_8$ tension and its Gaussian process forecasting}

\author{Mauricio Reyes}
\email{mauricio.cruzreyes@epfl.ch}
\affiliation{Institute of Physics, Laboratory of Astrophysics, EPFL, 1290 Versoix, Switzerland.}

\author{Celia Escamilla-Rivera}
\email{celia.escamilla@nucleares.unam.mx}
\affiliation{Instituto de Ciencias Nucleares, Universidad Nacional Aut\'onoma de M\'exico, 
Circuito Exterior C.U., A.P. 70-543, M\'exico D.F. 04510, M\'exico.}


\begin{abstract}

In this paper we reconstruct the growth and evolution of the cosmic structure of the Universe using Markov Chain Monte Carlo algorithms for Gaussian processes \cite{titsias2008markov}. We estimate the difference between the reconstructions that are calculated through a maximization of the kernel hyperparameters and those that are obtained with a complete exploration of the parameter space. We find that the difference between these two approaches is of the order of $1\%$. Furthermore, we compare our results with those obtained by Planck Collaboration 2018 assuming a $\Lambda$CDM model and we do not find a statistically significant difference in the redshift range were the reconstructions of $f\sigma_{8}$  have been made.
\end{abstract}

\maketitle


\section{Introduction}

Currently, the estimates of the value of $ S_ {8}\equiv \sigma_{8} \sqrt{\Omega_{m}/0.3} $, obtained from the $\Lambda$CDM fit to the  Cosmic Microwave Background (CMB)  differ between 2-3~$\sigma$ with respect the value of $S_{8}$ obtained with the analysis of galaxy clustering using    two-point correlation functions (2PCFs) \cite {DES:2021wwk}. This discrepancy is the so-called $S_{8}$ tension  \cite{Troster:2019ean,DiValentino:2020vvd}.  Additionally, the most recent estimate of the value of the Hubble constant, $H_{0}$, obtained with the calibration of the cosmic distance ladder scale through Cepheid stars and supernovae type Ia  is $5 \sigma$ different \cite{riess2021comprehensive} from  the value $\Lambda$CDM obtained with CMB observations.

In specific, the primary anisotropies of the CMB exhibit a tension in the matter clustering strength at the level of $2-3\sigma$ when compared to lower $z$ probes such as weak gravitational lensing and galaxy clustering (e.g.~\cite{Asgari:2019fkq,KiDS:2020suj,Joudaki:2019pmv, DES:2021wwk,DES:2021bvc,DES:2021vln,KiDS:2021opn,  Hildebrandt:2018yau,DES:2020ahh,Philcox:2021kcw}). 
 The lower $z$ probes (see Figure 4 from \cite{Abdalla:2022yfr}), select a lower value of $S_8$ compared to the high $z$ CMB estimates. The measured $S_8$ value is model dependent and in the majority of the scenarios it is considered a standard flat $\Lambda$CDM model. Of course, this latter model provides a well fit to the data from all probes, but predicts a lower value of structure formation compared to what we expect from the CMB \cite{KiDS:2020suj}. In \cite{Abdalla:2022yfr}  it is reported latter
 parameter estimates and constraints, where we notice that there might be slightly differences  in the selection that enter each studies. For example, on one hand a statistical property of the $S_8$  distribution might be selected, such as its mean or mode together with the asymmetric $68\%$ C.L around this value or the standard deviation of the data points. On the other hand,  the statistics to the full posterior distribution can be adopted, such as the maximum a posteriori point or the best fitting values and its errors. Anyhow, these considerations
 can affect the estimated values of the parameters, in particular when the posterior distributions are significantly non-Gaussian.
 
Many beyond $\Lambda$CDM models have been proposed as tentative explanations to this observed $S_8$ tension, however non specific model has been proven much better than the standard one. In such case, we should look for a reason in where the tension between CMB and cosmic shear in the inferred value of  $S_8$ can arise from unaccounted baryonic physics, other unknown systematics or a statistical fluke. Furthermore, we require that $S_8$ can be independent of CMB and cosmic shear. In this matter, redshift-space distortion (RSD) data also prefer a small value of  $S_8$ that is 2-3$\sigma$ lower than the Planck result \cite{Nunes:2021ipq}. However, the RSD values on $S_8$ is sensitive to the cosmological model.

In this \textit{paper} we perform the reconstructions of the   $f\sigma_{8}(z)$ observations using Gaussian processes (GP) to analyze if the reconstructions suggest possible deviations with respect the standard cosmological model. Unlike previous studies  \cite{Benisty:2020kdt,Li:2019nux,LeviSaid:2021yat}, we take into account the Planck 2018 confidence contours when comparing the predictions of $\Lambda$CDM with those of the reconstructions. This allows to keep in control the statistical uncertainties.
The possibility of using GP to distinguish between Modified Gravity (MG) and General Relativity (GR) has been analyzed \cite{Reyes:2021owe}, e.g. treating the perturbations of a disformally scalar field model which background mimics the $\Lambda$CDM \cite{Dusoye:2021jne}. 

Previous articles \cite{Li:2019nux} have used the reconstructions of the Hubble parameter to reconstruct $f\sigma_{8}(z)$ and to show how different values of   $H_{0}$ change the  value of the  $f\sigma_{8}$ tension.  
However, the large number of free parameters associated with that approach leads to large uncertainties on the reconstructions, which could explain why a $\sim 4\sigma$ change in the value of $H_{0}$ only  causes a $\sim 1\sigma$ change   on the  value of the $f\sigma_{8}$ tension. 

Unlike the standard way to reconstruct $f\sigma_8$ \cite{Li:2019nux}, in this work we reconstruct $f\sigma_{8}$(z) using directly the  estimations  of $f\sigma_{8}$, therefore it will be possible to remove  $H_{0}$ and $\sigma_{8}$ as free parameters and reduce the uncertainty in the confidence contours. However, we should mention that   our approach  does not allow us to obtain a direct estimate of the aforementioned parameters.

In this line of thought, several numerical methods have been developing upon these years, showing a great advance in the precision cosmology road, e.g. methods that involves artificial neural network (ANN) to reconstruct late–time cosmology data \cite{Dialektopoulos:2021wde}, creation of mock datasets through machine learning (ML)  based on the LSST survey and using a fiducial cosmology \cite{Arjona:2021mzf} and bayesian analyses in order to re-assessed the $\sigma_8$ discrepancy between CMB and weak lensing data \cite{Nunes:2021ipq}.

Furthermore, we analyse two different approaches to obtain the value of the confidence contours of  the reconstructions. The first approach consists on the maximization of the likelihood associated with the GP in order to obtain the value of the free parameters from the reconstruction. This approach has been frequently used in the literature to perform GP in the context of cosmology \cite{Briffa:2020qli,Reyes:2021owe,Li:2019nux,Perenon:2021uom,Ruiz-Zapatero:2022zpx}, however in some works \cite{seikel2012reconstruction,Escamilla-Rivera:2021rbe}  it has 
 been shown that this process could lead to an underestimation of the confidence contours of the reconstructions. The second approach  consists in an exploration of the parameter space of the free parameters through Markov Chains Monte Carlo  (MCMC) methods. This  method allows us to obtain an estimate of the uncertainties associated with each parameter, which contributes to solve the underestimation problem.

\section{Treatment of  data samples and methodology}  

To use GP method, we need to assume that our set of observations ``$\textbf{y}$" given in the set of redshifts ``$\boldsymbol{z}$" is Gaussian distributed around the underlying function that we seek to reconstruct, $\textbf{g(z)}$. This allows us to associate a probability distribution with the data:
\begin{equation}
    \textbf{y}\sim \mathcal{N}(\overline{\mu}, K(\textbf{z},\textbf{z})+C),\label{eq:dis1}
\end{equation}
where $\overline{\mu} $ is the mean value of the observations, $ C $ is its  covariance matrix  and $ K $ is the covariance matrix associated with the Gaussian processes also known as  kernel function. It can be shown that the mean value and covariance matrix of the reconstructed function in the set of redshifts $\boldsymbol{z}^{*}$ is given by \cite{seikel2012reconstruction} 
\setlength{\belowdisplayskip}{0pt} 
\begin{gather}
\overline{g}=\overline{\mu}^{*}+K(\boldsymbol{z}^{*},\boldsymbol{z})\left [ K(\boldsymbol{z},\boldsymbol{z})+C \right ]^{-1}(y-\overline{\mu}),  \label{eq:48}
\end{gather}
\begin{equation}
    C(g^{*})=K(\boldsymbol{z}^{*},\boldsymbol{z}^{*})-K(\boldsymbol{z}^{*},\boldsymbol{z})\left [ K(\boldsymbol{z},\boldsymbol{z})+C \right ]^{-1}K(\boldsymbol{z},\boldsymbol{z}^{*}). 
\end{equation}

Measurements of the clustering pattern of matter in the redshift space  allows us to infer  parameters such as the linear  growth of matter perturbations $f(z)$ and the variance of matter fluctuations $\sigma_{R}^{2}$ on a given scale $R$. Variance  is usually reported on a scale  $R=8 h^{-1}$ Mpc, that is, $\sigma_{8}^{2}$. 

However, an exact determination of the value of $f(z)$ turns out to be complex, since the galaxy redshift surveys  provide an estimate of the perturbations in terms of galaxy densities  $\delta_{g}$ instead  not directly in terms of matter density   $\delta_{g}=b \delta_{m}$. 
Unfortunately, the exact value of  the parameter $b$ remains uncertain \cite{nesseris2017tension}, for that reason, the measurements are reported in  terms of $f\sigma_{8}(z)$  since it is independent of the parameter $b$ \cite{Li:2019nux}. 

To perform the reconstructions we use the compilation of measurements of  $f\sigma_{8}(z)$   shown in Table I of \cite{perenon2019optimising}. This table contains 30 measurements of $f\sigma_{8}$, together with their uncertainties and their covariance matrices.   

According to the latter, the inferred values of $f\sigma_{8}(z)$ depend on two things: \textit{(i)} the  anisotropies in the power spectrum of peculiar velocities of galaxies and \textit{(ii)} the fiducial cosmology chosen to estimate the values of  $f\sigma_{8}(z)$.   If the cosmology chosen to perform the data analysis does not adequately describe the geometry of the universe, then nontrivial  anisotropies are introduced in the 2PCFs, which are directly correlated with the estimated value of $f\sigma_{8}$, this effect is known as Alcock-Paczynski  (AP) effect.  Is important to  that the $f\sigma_{8}(z)$ values sometimes are reported assuming  \cite{Kazantzidis:2018rnb} different fiducial cosmologies, e.g.
\begin{itemize}
    \item When considering a fiducial cosmology with $\Omega^{0}_{m} =  0.26479 $, $H_{0} = 71$~km/s/Mpc, $\sigma_{8} = 0.8$, $z_{\text{eff}} = 1.52$, it is obtained $f\sigma_{8}(z_{\text{eff}})= 0.420 \pm 0.076$ \cite{gil2018clustering}.
    \item While with a fiducial cosmology with $\Omega^{0}_{m} =  0.31 $, $H_{0} =67.6$~km/s/Mpc, $\sigma_{8} = 0.8225$, $z_{\text{eff}} = 1.52$ , we can obtain $f\sigma_{8}(z_{\text{eff}})= 0.396 \pm 0.079$  \cite{hou2018clustering}.
\end{itemize}


\section{Alcock-Paczynski corrections}
 
 If the $f\sigma_{8}(z)$ reconstructions are performed without including corrections to the AP effect, then all the information from different fiducial cosmologies would be mixed, i.e it cannot be possible to properly estimate the tension between the reconstructions and the $\Lambda$CDM model. We apply the correction to the AP effect given in \cite{Kazantzidis:2018rnb}, whose corrections state that if a measurement of $\overline{f\sigma_{8}}(z)$ has been obtained assuming a fiducial cosmology with a Hubble parameter $\overline{H}(z)$ and  angular diameter $\overline{D}_{A}(z)$, then the corresponding value of $f\sigma_{8}(z)$ assuming a different fiducial cosmology with $H(z)$ and $\overline{D}_{a}(z)$ can be approximated as: 
\begin{equation}
f\sigma_{8}(z)\approx \frac{H(z)D_{A}(z)}{\overline{H}(z)\overline{D}_{A}(z)}\overline{f\sigma}_{8}(z).\label{eq:correc}
\end{equation}
To correct the AP effect, we consider  a \textit{vanilla} $\Lambda$CDM cosmology    with the parameters inferred from Planck 2018 Collaboration \cite{Aghanim:2018eyx}, such as the high redshift data from Planck TT,TE,EE+lowE is $S_8=0.834\pm0.016$. Combining this data with secondary CMB anisotropies, in the form of CMB lensing, serves to tighten the constraint to $S_8=0.832\pm0.013$.

To proceed with this calculation, we set the following steps:
\begin{enumerate}
\item Suppose that $A(z_{i})$ and $B(z_{j})$ are two  measurements of $f\sigma_{8}(z)$ that were obtained assuming the fiducial cosmology $H(z)$
\begin{align}
A(z_{i})=\mu_{a}\pm\sigma_{a},    \\
B(z_{j})=\mu_{b}\pm\sigma_{b},    
\end{align}
with $\mu_{a}$, $\mu_{b}$  the mean value of each measurement and $\sigma_{a}$, $\sigma_ {b}$  their 1$\sigma$ uncertainties, also suppose that the measurements are correlated through a covariance matrix $C$ 
\begin{align}
C=\begin{pmatrix}
\sigma_{a}^{2} &  \sigma_{ab} \\ 
\sigma_{ba} & \sigma_{b}^{2}
\end{pmatrix}.
\end{align}
\item In order to compute the value of $A(z_{i})$ and $B(z_{j})$ in a  fiducial cosmology $H'(z)$,  we have to perform the AP correction. To do so, we multiply $A$ and  $B$  with  constants $C_{A}$ and $C_{B}$ given by the equation (\ref{eq:correc}) and then the covariance matrix for $AC_ {a}$ and $BC_ {b}$, is  given by \cite{wackerly2010estadistica}
\begin{equation}
C'=   \begin{pmatrix}
(C_{a}\sigma_{a})^{2} &  C_{a}C_{b}\sigma_{ab} \\ 
C_{a}C_{b}\sigma_{ba} & (C_{b}\sigma_{b})^{2}
\end{pmatrix}.
\end{equation}

\end{enumerate}


\section{Kernel metrics}

It has been shown that the kernel chosen to perform the reconstructions can affect the uncertainties of the parameters derived from the reconstructions \cite{OColgain:2021pyh,Escamilla-Rivera:2021rbe,seikel2012reconstruction}. In particular, the Gaussian kernel can lead to uncertainties up to three times smaller than the value of the uncertainties obtained with the Matérn covariance functions \cite{seikel2013optimising}, since this could be associated with the problem of underestimation of uncertainties, we choose to use only Matérn kernels, specifically the Matérn 3/2 and Matérn 5/2 kernels. These covariance functions have been used, e.g. in studies that analyze how the choice between different kernels affect the value of $H_{0}$ that is obtained from the reconstructions of the cosmic late expansion \cite{OColgain:2021pyh} and in analysis on how different kernels can affect the constraints on modified theories of gravity \cite{Briffa:2020qli}.  
The Matérn 3/2 and Matérn 5/2 kernels are respectively defined  as:
\begin{widetext}
\begin{eqnarray}
 \textbf{K}_{ij}= \eta^{2}\left ( 1+\frac{\sqrt{3(z_{i}-z_{j})^{2}}}{l} \right )\exp \left ( -\frac{\sqrt{3(z_{i}-z_{j})^{2}}}{l} \right ) \quad \rightarrow \quad \text{Matérn 3/2},\label{eq:Matern32} \\
 \textbf{K}_{ij}= \eta^{2}\left ( 1+\frac{\sqrt{5(z_{i}-z_{j})^{2}}}{l}+ \frac{5(z_{i}-z_{j})^{2}}{3l^{2}} \right )\exp \left ( -\frac{\sqrt{5(z_{i}-z_{j})^{2}}}{l} \right )  \quad \rightarrow \text{Matérn 5/2}, \label{eq:Matern52}
\end{eqnarray}
\end{widetext}
where $\eta$ and $l$ are free parameters that measure the width of the reconstructed function and the correlation between the function evaluated at two given points $z_{i}$ and $z_{j}$. With these kernels at hand we are ready to proceed with 
two different methods to obtain the value of  the hyperparameters of the kernel: 
\begin{itemize}
\item \textit{Method (i)} consists on the maximization of the likelihood associated with the observations Eq.~(\ref{eq:dis1}).   
\item \textit{Method (ii)} consists in a full exploration of  parameter space for the hyperparameters through MCMC methods. This approach is convenient when the parameter space is multidimensional  and it could help us to find the true maximum likelihood estimate in cases were the algorithm for maximization gets stuck in a local minima.  
\end{itemize}

\begin{figure*}
    \centering
\includegraphics[scale=0.17]{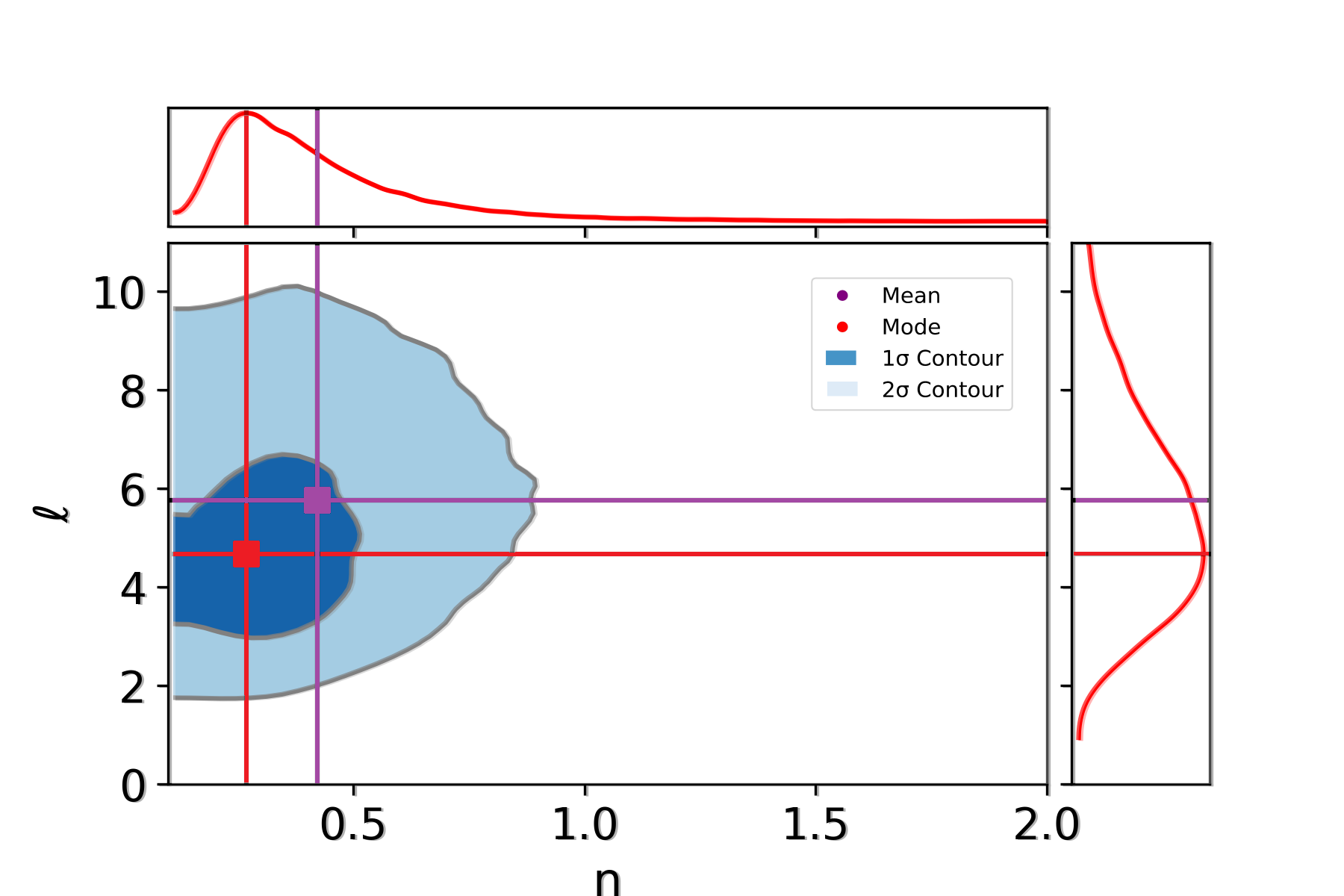}
\includegraphics[scale=0.17]{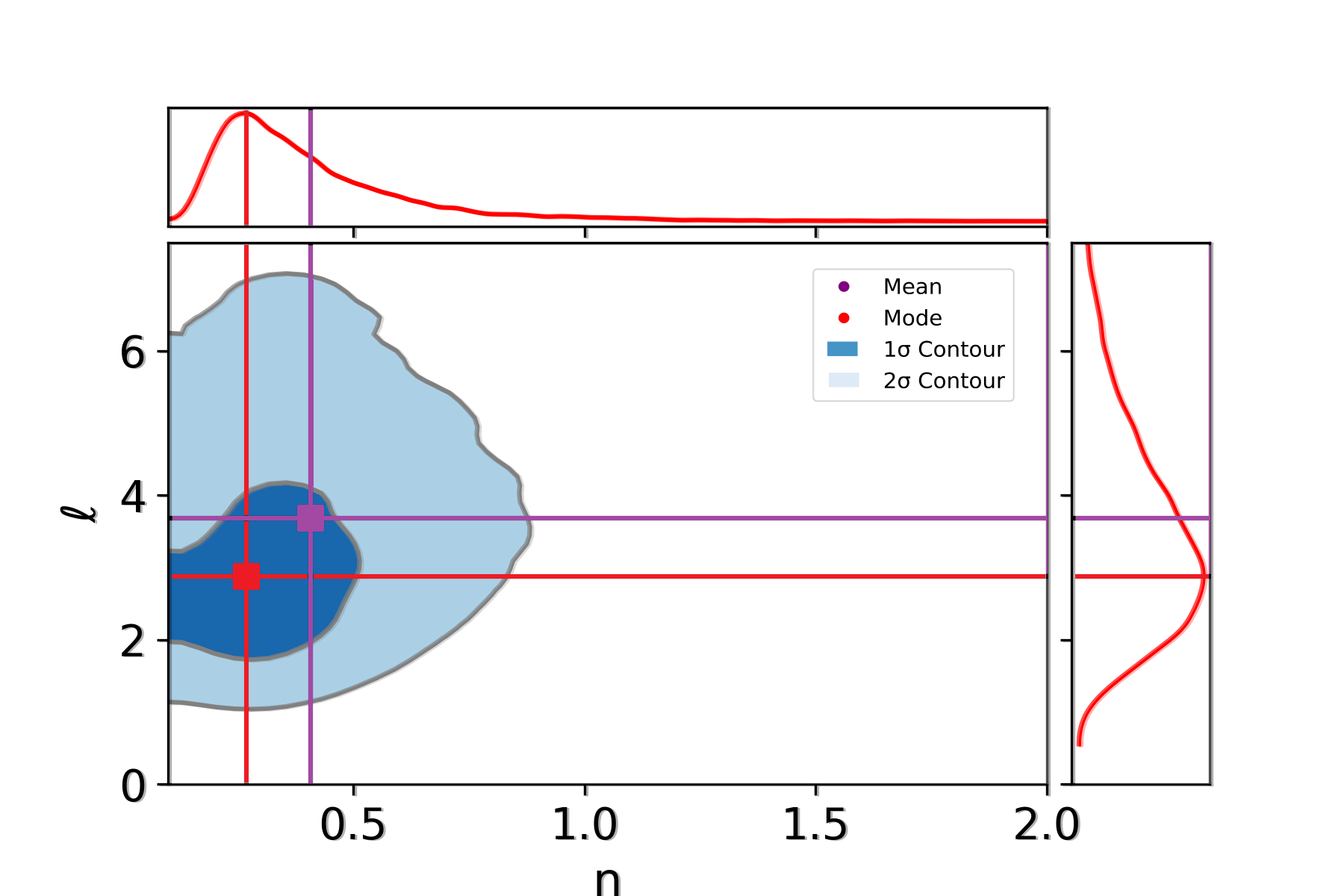}
\caption{\textit{Left:} 1$\sigma$ and 2$\sigma$ confidence contours for the hyperparameters of the Matérn 3/2 kernel (\ref{eq:Matern32}).
\textit{Right:} 1$\sigma$ and 2$\sigma$ confidence contours for the hyperparameters of the Matérn 5/2 kernel (\ref{eq:Matern52}). The posterior that appear on top and right side show the marginal distributions for each hyperparameter. The purple and red color dots show the region of parameter space where the mean (purple)  and mode (red) of each hyperparameter intersects. 
}
    \label{fig:contornos32}
\end{figure*}

Since $\eta$ is a squared quantity in both covariance functions  (\ref{eq:Matern32})-(\ref{eq:Matern52}), if  $(\eta_{\text{Max}},l_{\text{Max}})$ are the values of the  hyperparameters that maximize Eq.(\ref{eq:dis1}), then  $(-\eta_{\text{Max}},l_{\text{Max}})$  also
maximize Eq.(\ref{eq:dis1}), this leads to a bimodal posterior distribution for the hyperparameter $\eta^{M}$. Since with one mode we can obtain the full information to carry out the reconstructions, it is convenient to establish priors that only take into account the positive (or negative) branch  of the parameter space for $\eta$.  With this method it is possible to reduce the computational time required to calculate the posterior distribution of the hyperparameters and also it allow us to avoid convergence problems within the numerical code.

\begin{figure*}
    \centering
\includegraphics[scale=0.19]{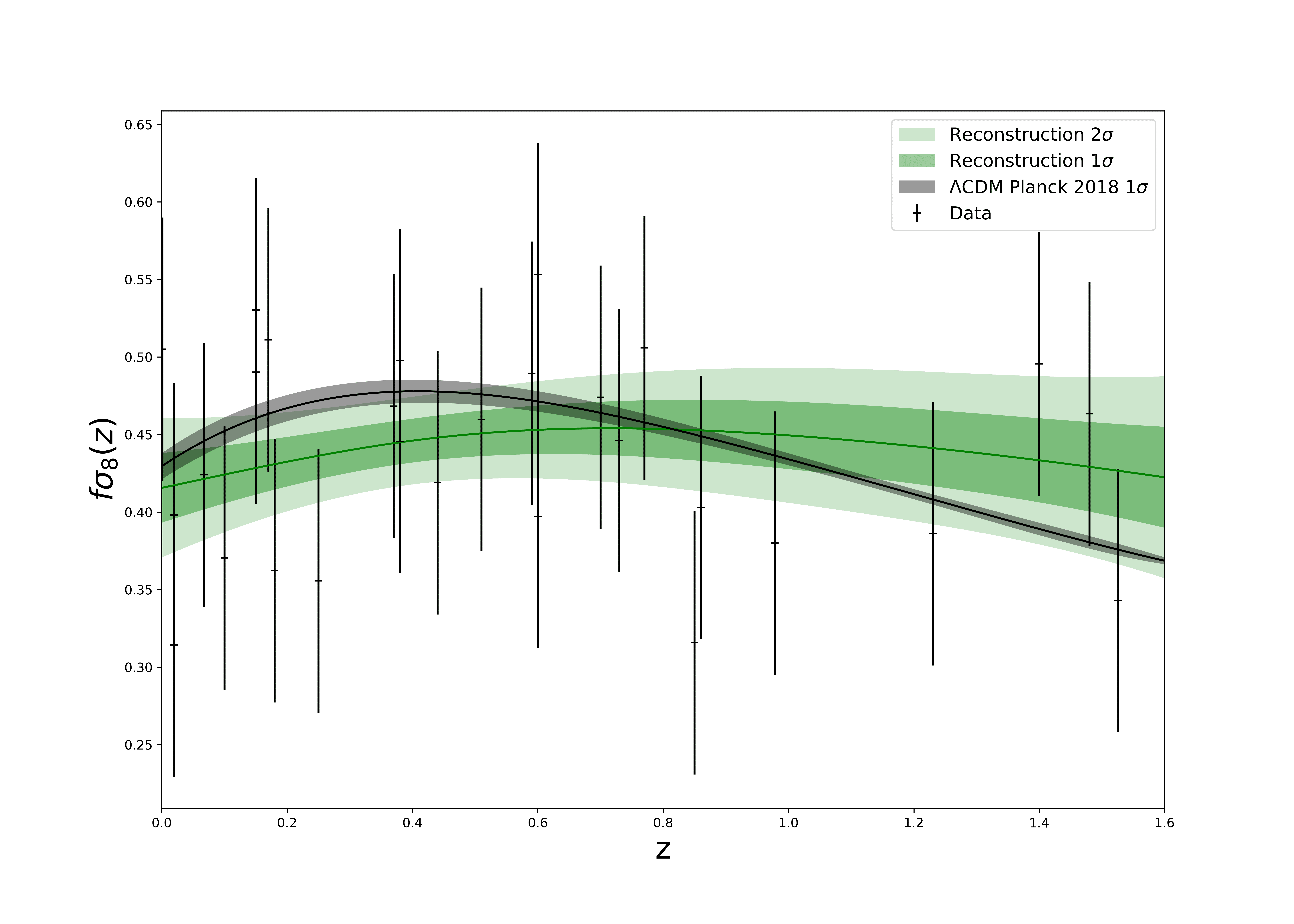}
\includegraphics[scale=0.19]{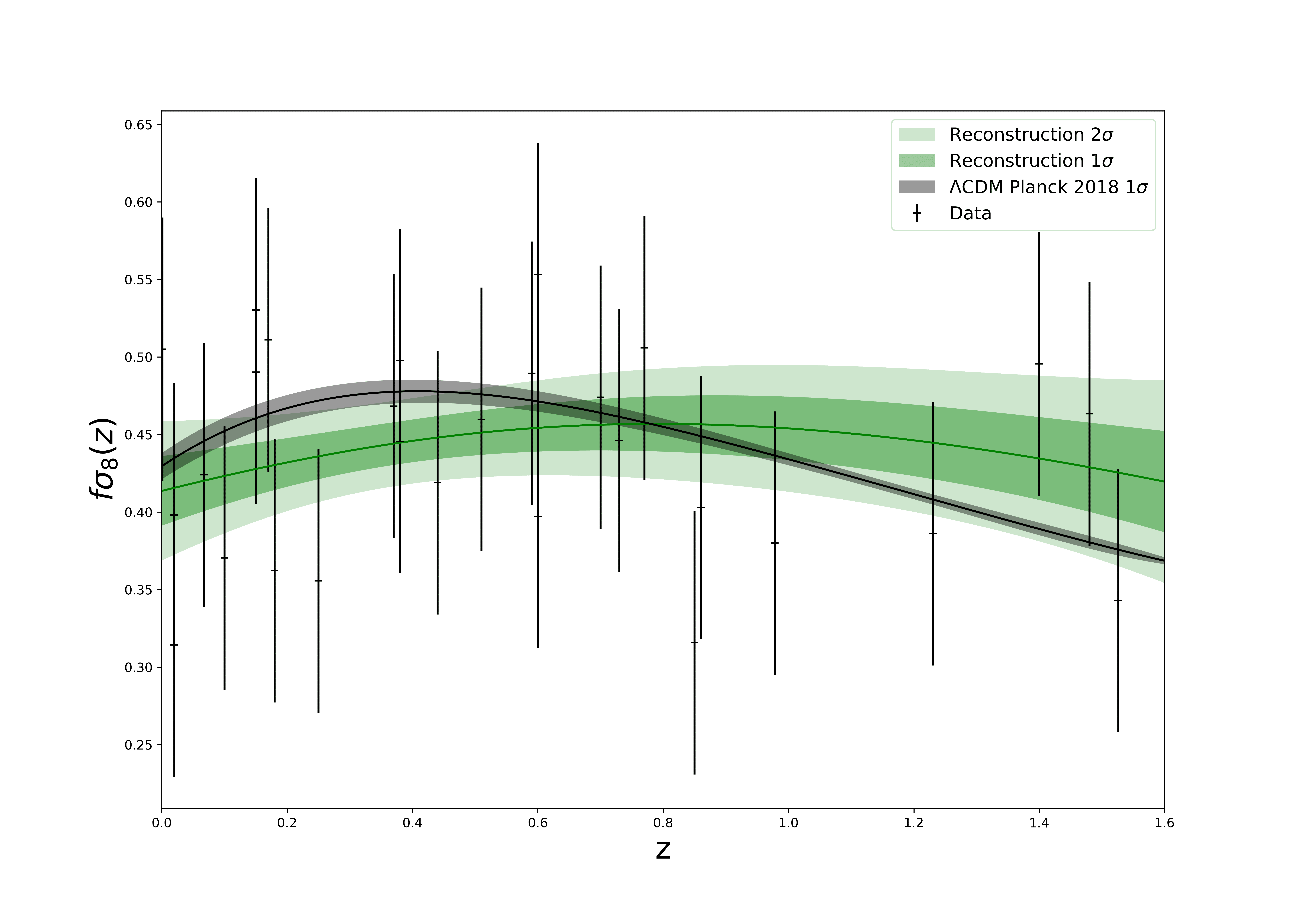}
\includegraphics[scale=0.19]{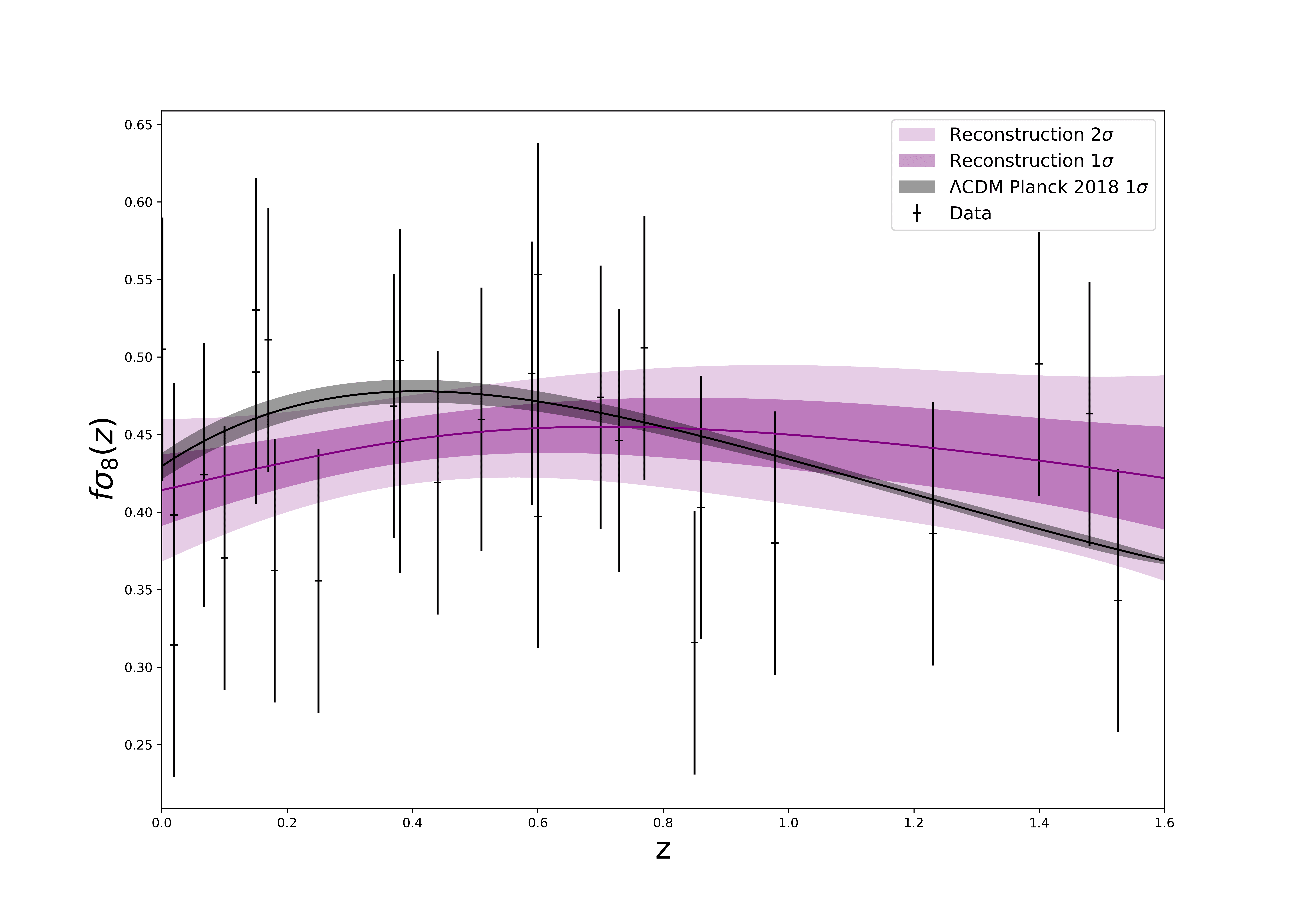} 
\includegraphics[scale=0.19]{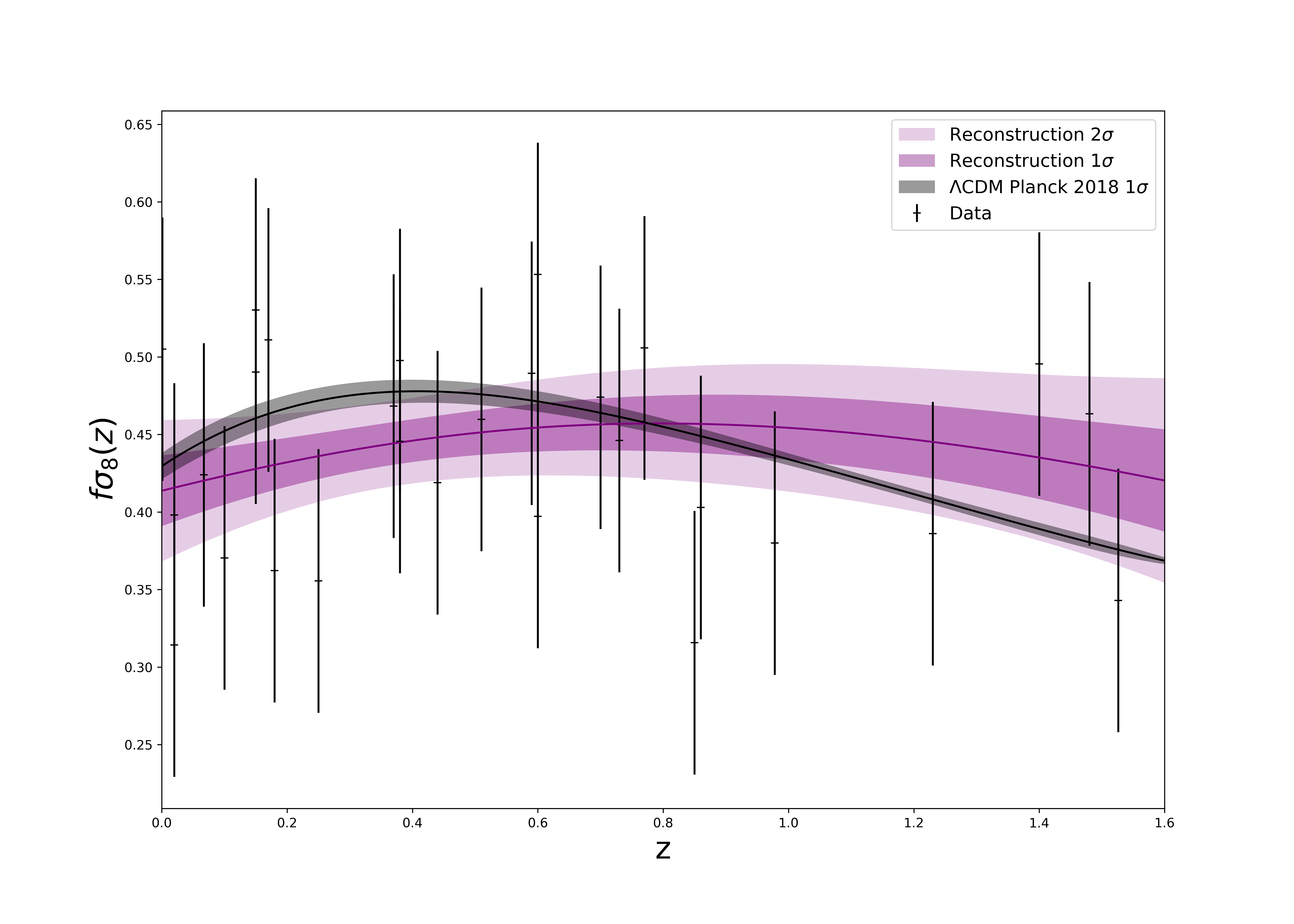}
\includegraphics[scale=0.19]{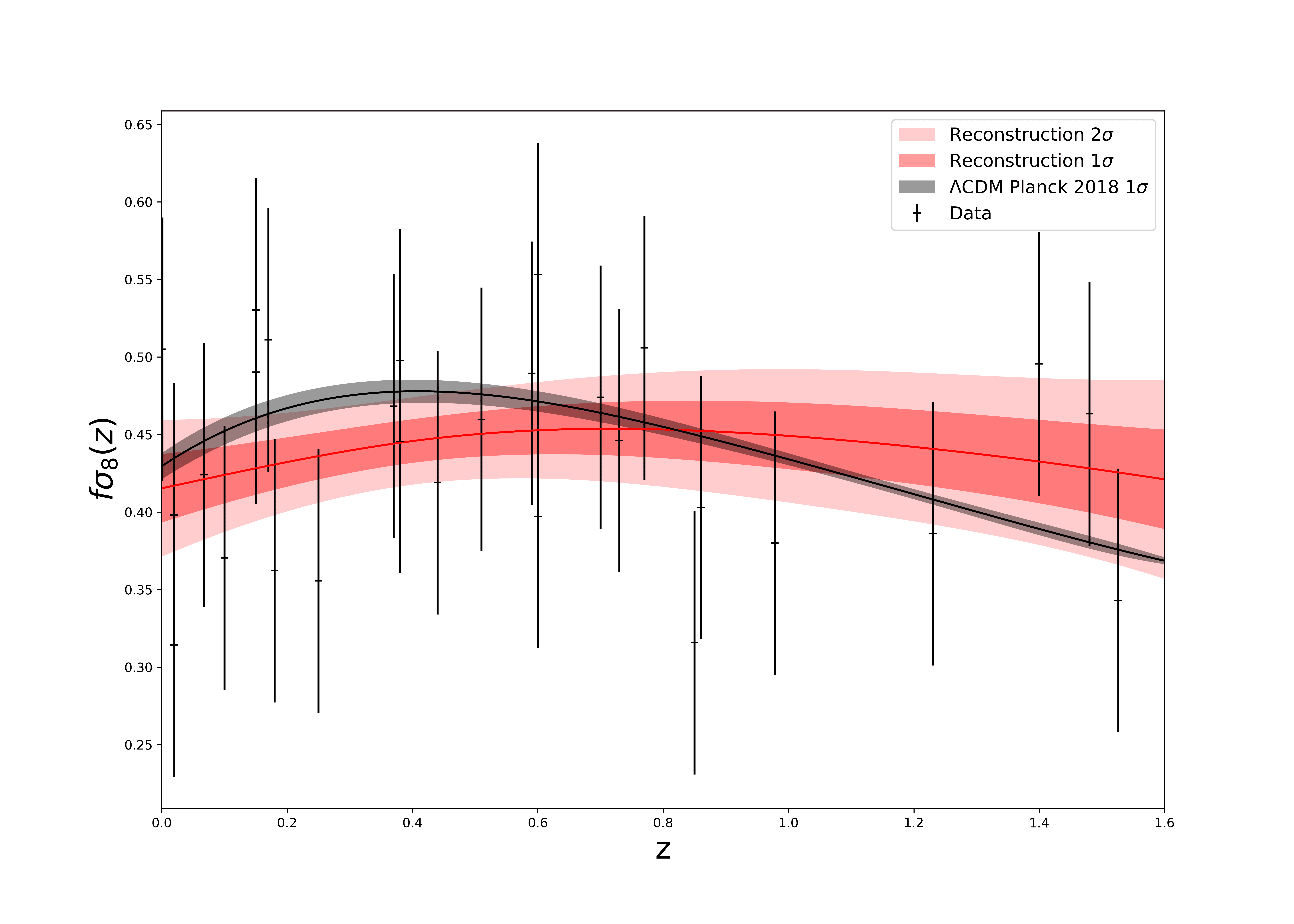}
\includegraphics[scale=0.19]{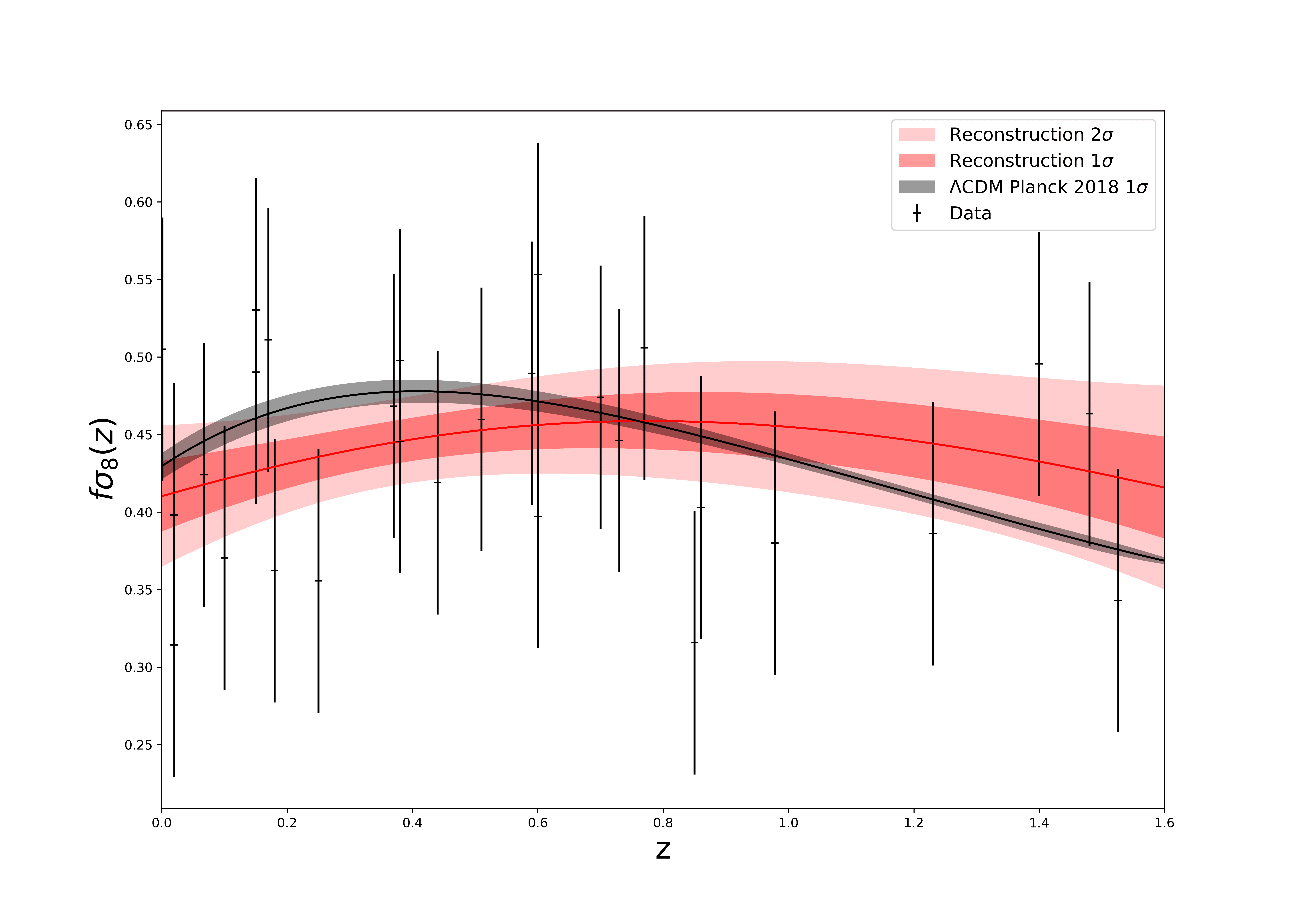}  
    \caption{\textit{Left:} Reconstructions of $f\sigma_ {8}(z)$
    with the Matérn 3/2 kernel (\ref{eq:Matern32}). \textit{Right:}
    Reconstructions of $f\sigma_ {8}(z)$ with the Matérn 5/2 kernel (\ref{eq:Matern32}). 
  \textit{Top}: Reconstructions obtained using the Maximum likelihood method (green color). 
  \textit{Middle:} Reconstructions obtained using the mean value of the hyperparameters (purple color). 
  \textit{Bottom:} Reconstructions obtained using the mode of the hyperparameters (red color).  The data is denoted by the black points.
  In gray we shown the prediction for $f\sigma_ {8} (z)$ given a fitted
     $\Lambda$CDM with CMB Planck 2018 values \cite{Aghanim:2018eyx}. Confidence contours of each reconstruction are shown at   1$\sigma$ and 2$\sigma$ level.} %
    \label{fig:rec32}
\end{figure*}

Notice that the covariance functions (\ref{eq:Matern32}) and (\ref{eq:Matern52}) are not symmetric on $l$. However, we expect positive values of $l$, otherwise the correlation between the points $z_ {i}$, $z_{j}$    will grow proportionally to
\begin{align}
  \textbf{K}_{ij} \propto
  \exp \left ( -\frac{ \left | z_{i}-z_{j} \right |}{l} \right ).\label{eq:proportionally}
\end{align}
Since hyperparameters are restricted to be positive,  we consider for them Gamma probability distributions as priors.  The  probability density function $P(X)$ for a random variable $X$ that follows a Gamma distribution with parameters  $\alpha$ and $\beta$, is given by
\begin{align}
  P(X)=
  \begin{cases}
   \hfil 0    & \text{if } X \leq 0, \\
    \hfil \frac{    \beta (\beta X)^{\alpha-1}    e^{-\beta X} }{\Gamma(\alpha)}                
    & \text{if } X>0,
\end{cases}
\end{align}
with $\Gamma$  the Gamma function. The mean value  of the $P(X)$  is  $M =\alpha/\beta$, the mode is $M_{0}= (\alpha-1)/\beta$ and the variance is Var$(X^{2}) = \alpha/\beta^{2}$ \cite{wackerly2010estadistica}. In order to find the value of the hyperparameters of the covariance functions that maximizes Eq.(\ref{eq:dis1}) we perform a Maximum Likelihood Estimation (MLS) with the code \cite{seikel2012reconstruction}, and afterwards we proceed with a standard MCMC analysis using the public available code PyMC3 \cite{salvatier2016probabilistic}. For both hyperparameters we choose Gamma priors with $\beta = 1$  and   $\alpha$ equal to their MLS. We estimate the convergence of the chains using a Gelman-Rubin convergence criteria \cite{Gelman:1992zz}  with $R - 1 < 0.03$.  The reconstructions are estimated with two different approaches: (1) we use the mean value of the hyperparameters to estimate the mean value of the reconstructions and (2) we use the mode.  Our  results are  detailed in  Tables \ref{table:summary32}-\ref{table:summary52} and in Figure \ref{fig:rec32}. 

\begin{table*}
\centering
\begin{tabular}{ |c  | c|c |c|c |}
\hline
Parameter     & MLS & Mean value & Standard deviation & Mode \\
\hline
$l$  & 5.36  &  5.73   & 2.25 &  4.50\\
$\eta$ &  0.33  &    0.42 & 0.24  &   0.26 \\
\hline
\end{tabular}
\caption{Hyperparameters statistics summary for the Matérn 3/2 kernel (\ref{eq:Matern32}). \textit{First column:} indicates the kernel hyperparameters. \textit{Second column:} denotes the value  Maximum Likelihood Estimation (MLS) for the hyperparameters. \textit{Third column:} shows the mean value of the posterior distribution for each hyperparameter obtained using a MCMC. \textit{Fourth column:} shows the standard deviation. \textit{Fifth column:}indicates the mode.}
\label{table:summary32}
\vspace{2\baselineskip}
\centering
\begin{tabular}{ |c  | c|c |c|c |}
\hline
Parameter     & MLS & Mean value & Standard deviation & Mode \\
\hline
$l$  &  3.33 &    3.71 & 1.70 & 2.80 \\
$\eta$ &  0.32  &    0.41 & 0.26  & 0.27    \\
\hline
\end{tabular}
\caption{Hyperparameters statistics summary for the Matérn 5/2 kernel (\ref{eq:Matern52}). The meaning of columns is the same as in Table \ref{table:summary32}.}
\label{table:summary52}
\end{table*}

\begin{table*}
\centering
\begin{tabular}{|c|c|c|c|}
\hline
Method     & Redshift of maximum difference & Maximum difference & Total difference  \\
\hline
Mean  &  0.30 & 2.09 & 1.95    \\
Mode  & 0.31 &   2.15  & 2.00 \\
MLS  &  0.31 &  2.13 & 1.99   \\
\hline
\end{tabular}
\caption{Comparison between the different methods analyzed in this work and $\Lambda$CDM. First column shows the method used to  estimate the value of the hyperparameters of  the Matérn 3/2 kernel. Second column shows the redshift where Eq.(\ref{eq:tension}) reach his maximum, third column shows the value of $T(z)$ reached at that redshift and the fourth  shows the result of integrating $T(z)$ in the redshift range were the reconstructions were performed. }
\label{table:difference32}
\vspace{2\baselineskip}
\centering
\begin{tabular}{|c|c|c|c|}
\hline
Method     & Redshift of maximum difference & Maximum difference & Total difference  \\
\hline
Mean  & 0.3  &  2.18   & 2.12  \\
Mode  & 0.3   &  2.17   & 2.06 \\
MLS  &  0.3  &  2.19  &  2.12 \\
\hline
\end{tabular}
\caption{Comparison between the different methods analyzed in this work and $\Lambda$CDM. The description by columns are the same as the ones reported in Table \ref{table:difference32} but for the Matérn 5/2 kernel.}
\label{table:difference52}
\end{table*}


In order to distinguish between reconstructions we define the tension metric as follows: 
\begin{equation}
    T(z) = \frac{|f\sigma_{8}^{rec}(z) - f\sigma_{8}^{\Lambda\text{CDM}}(z)| }{\sqrt{\sigma_{\text{rec}}^{2}(z)+\sigma_{\Lambda\text{CDM}}^{2}(z) }},\label{eq:tension}
\end{equation}
where $f\sigma_{8}^{rec}(z)$, $f\sigma_{8}^{\Lambda\text{CDM}}(z)$ are the mean values of the reconstructions and $ \Lambda\text {CDM }$, respectively. $\sigma_{\text{rec}(z)} $, $\sigma_{\Lambda \text {CDM}(z)} $ are the $1\sigma$ statistical uncertainties.  Notice that Eq.(\ref{eq:tension}) quantifies the difference in standard deviations between
the reconstructions and  $ \Lambda\text{CDM} $. 

From Tables \ref{table:difference32} and \ref{table:difference52} we can notice that for  both kernels  the tension for all methods is of the order of  2$\sigma$ in the observable redshift region.  The mean value of the reconstructions obtained through the different methods represent $1\%$ of its total value.

Additionally, to test the performance of each model describing by the observations, we estimate  the chi-square statistics as follows
\begin{align}
\Delta f\sigma_{8} \mid_{i} = f\sigma_{8_{\text{obs}}}(z_{i}) - f\sigma_{8_{\text{recons}}}(z_{i}),\\
    \chi^{2} =  \Delta f\sigma_{8}^{T} \cdot C^{-1} \cdot \Delta f\sigma_{8},
\end{align}
with $C^{-1}$ the inverse of the covariance matrix of the data. The results are shown in Tables \ref{table:chi32} and \ref{table:chi52}, with these results and with those obtained using the $T(z)$ function, we conclude that there is no statistical significant difference between $\Lambda$CDM and the reconstructions presented here (see Tables \ref{table:chi32} and \ref{table:chi52}). 
\begin{table*}
\centering
\begin{tabular}{|c|c|}
\hline
Method     & $\chi^{2}$ \\
\hline
$\Lambda$CDM &   0.85 \\
Mean  &   0.77\\
Mode  &  0.78\\
MLS &   0.77 \\
\hline
\end{tabular}
\caption{$\chi^{2}$-statistics over the number of degrees of freedom ($N = 28$) for $\Lambda$CDM and the different reconstructions obtained using the  kernel Matérn 3/2.}
\label{table:chi32}
\vspace{2\baselineskip}
\centering
\begin{tabular}{|c|c|}
\hline
Method     & $\chi^{2}$ \\
\hline
Mean  &  0.78      \\
Mode  &  0.77\\
MLS &    0.78 \\
\hline
\end{tabular}
\caption{$\chi^{2}$-statistics over the number of degrees of freedom ($N = 28$) for the different reconstructions obtained using the  kernel Matérn  5/2.}
\label{table:chi52}
\end{table*}


\section{Conclusions} 

In this paper we reconstructed the $f\sigma_{8}$ observations using two different kernels and three different methods to obtain the value of the hyperparameters. 
The tension between $\Lambda$CDM and the reconstructions does not give values
above 2.2$\sigma$. We showed that choosing different kernels leads to  differences of the order of $\sim 0.1\sigma$ in the value of the tension. Furthermore, we also show that the change in the value of the tension obtained with different methods (
with the kernel fixed) is also of the order  of $ 0.1\sigma$. Therefore,  we conclude that there is not significant statistical difference between the predictions given for $f\sigma_{8}(z)$ by  $\Lambda$CDM with Planck 2018 parameters and those given by the reconstructions, since the maximum tension between both is below $2.2\sigma$.


\textit{Acknowledgments.-}
MC  acknowledge support from the European Research Council (ERC) under the European Union's Horizon 2020 research and innovation programme (Grant Agreement No. 947660). 
CE-R is supported by DGAPA-PAPIIT UNAM Project TA100122 and acknowledges the Royal Astronomical Society as FRAS 10147.
This work is part of the Cosmostatistics National Group (\href{https://www.nucleares.unam.mx/CosmoNag/index.html}{CosmoNag}) project. The Authors would
like to acknowledge the  PyMC3 and Arviz communities for their helpful recommendations on the modified version of arviz code \cite{arviz_2019}. 

\bibliographystyle{unsrt}
\bibliography{refs}

\end{document}